\documentclass[11pt]{article}

\advance \textheight by \topskip
\makeatletter

\def\vp{\vec{p}}

\def\vv{\vec{v}}

\usepackage{amsmath,amssymb}

\begin{document}

\title{
{\bf Anyons and the Landau problem in the noncommutative plane}}

\author
{{\sf Mikhail S. Plyushchay}\thanks{E-mail: mplyushc@lauca.usach.cl}
\\[4pt]
 {\small \it Departamento de F\'{\i}sica,
Universidad de Santiago de Chile}\\
{\small \it Casilla 307, Santiago 2, Chile}\\
}
\date{}

\maketitle

\begin{abstract}
The Landau problem in the noncommutative plane is discussed in the
context of  realizations of the two-fold centrally extended planar
Galilei group and the anyon theory.
\end{abstract}


\vskip 0.3cm\noindent


In 2+1 dimensions, Galilei group admits a two-fold central extension
\cite{LL,DH}
 characterized by the algebra with the nonzero Poisson bracket
relations
\begin{equation}\label{Galex1}
\{{\cal K}_i,{\cal P}_j\}=m\delta_{ij},\qquad \{{\cal K}_i,{\cal
K}_j\}=-\kappa\epsilon_{ij},
\end{equation}
\begin{equation}\label{Galex2}
\{{\cal K}_i,{\cal H}\}={\cal P}_i, \qquad \{{\cal J}, {\cal
P}_i\}=\epsilon_{ij}{\cal P}_j,\qquad
 \{{\cal J},{\cal
K}_i\}=\epsilon_{ij}{\cal K}_j,
\end{equation}
where $m$ and $\kappa$ are  the central charges. The algebra has the
two Casimir elements
\begin{equation}\label{Cas}
    {\cal C}_1=m{\cal J} +\kappa{\cal H}-\epsilon_{ij}{\cal
    K}_i{\cal P}_j,\qquad
    {\cal C}_2=m{\cal H}-\frac{1}{2}{\cal P}_i^2,
\end{equation}
which correspond to the (multiplied by the mass $m$) internal
angular momentum (spin) and energy.

There are two possibilities to realize this algebra as a symmetry of
a free particle on a plane: the \emph{minimal} realization and the
\emph{extended} one [cf. the two formulations for a free
relativistic anyon \cite{CP}].  Requiring that the particle
coordinate $X_i$  forms a Galilei covariant object with respect to
the action of the generators $J$, ${\cal P}_i$ and ${\cal K}_i$,
treating  the Galilei generators as integrals of motion and
identifying the  ${\cal P}_i$ as the canonical momentum  $p_i$, and,
finally,  putting the spin and internal energy to be equal to zero
(${\cal C}_1={\cal C}_2=0$), we arrive at the following realization
of the generators:
\begin{equation}\label{Min}
{\cal P}_i=p_i,\qquad K_i=mX_i-tp_i+m\theta\epsilon_{ij}p_j,\qquad
{\cal J}=\epsilon_{ij}X_ip_j+\frac{1}{2}\theta \vp\,{}^2,\qquad
{\cal H}=\frac{1}{2m}\vp\,{}^2,
\end{equation}
$\theta=\kappa/m^2$. As a result, the $X_i$ has a usual free
particle evolution, $\dot{X}_i=\frac{1}{m}p_i$. The price we pay for
such a minimal realization of the exotic Galilei algebra is the
non-commutativity of the coordinate components
\begin{equation}\label{XX}
    \{X_i,X_j\}=\theta\epsilon_{ij},
\end{equation}
and the non-canonical form of the
associated  symplectic structure
\begin{equation}\label{omega}
    \sigma_0=dp_i\wedge dX_i+\frac{1}{2}\theta\epsilon_{ij}dp_i\wedge
    dp_j.
\end{equation}

One can define another sort of the coordinate \cite{HPem,OP},
\begin{equation}\label{Y}
    Y_i=X_i+\theta\epsilon_{ij}p_j.
\end{equation}
It has the same bracket with $p_j$,
\begin{equation}\label{Yp}
    \{Y_i,p_j\}=\delta_{ij},
\end{equation}
 and, hence, the same evolution law as the coordinate $X_i$.
In terms of the $Y_i$ and $X_i$, the symplectic structure and
angular momentum are diagonal,
$$
\sigma_0=\frac{1}{2\theta}\epsilon_{ij}\left(dY_i\wedge dY_j -
dX_i\wedge dX_j\right),  \qquad {\cal
J}=\frac{1}{2\theta}\left(Y_i^2-X_i^2\right).
$$
On the other hand, in terms of the $Y_i$ and $p_i$  the boost generator
is represented in the usual form ${\cal K}_i=mY_i-tp_i$. However,
the $Y_i$, unlike the $X_i$, is not covariant with respect to the
Galilei boosts, $\{{\cal
K}_i,Y_j\}=t\delta_{ij}-m\theta\epsilon_{ij}$. As we shall see
below, the importance of the coordinate (\ref{Y}) reveals under
coupling the system to the external electric and magnetic fields.

Due to the noncommutative nature of the both $X_i$ and $Y_i$, there
is no coordinate representation associated with them. But since
\begin{equation}\label{YYX}
    \{Y_i,Y_j\}=-\theta \epsilon_{ij},\qquad \{X_i, Y_j\}=0,
\end{equation}
one can define the third sort of  the coordinate,
\begin{equation}\label{calX}
    {\cal X}_i=\frac{1}{2}(X_i+Y_i).
\end{equation}
It has commuting components and reduces the symplectic structure and
angular momentum to a canonical form,
$$
\sigma_0=dp_i\wedge d{\cal X}_i,\qquad {\cal J}=\epsilon_{ij}{\cal
X}_ip_j.
$$
Like the $Y_i$, the coordinate ${\cal X}_i$  is not covariant with
respect to the Galilean boosts, $\{{\cal K}_i,{\cal
X}_j\}=t\delta_{ij}-\frac{1}{2}m\theta\epsilon_{ij}$. The importance
of this third coordinate is that at the quantum level it provides us
with the Schr\"odinger representation,
 $\hat{\cal X}_i\Psi({\cal
X})={\cal X}_i\Psi({\cal X})$,
$\hat{p}_i=-i\partial_i\Psi({\cal X})$. In this
representation in accordance with Eqs. (\ref{calX}), (\ref{Y}) the
action of the covariant coordinate operator is reduced to the star
multiplication \cite{star}:
$$
\hat{X}_i\Psi({\cal X})=\left({\cal
X}_i-\frac{i}{2}\theta\epsilon_{ij}\partial_j\right)\Psi({\cal
X})\equiv {\cal X}_i\star \Psi({\cal X}).
$$

We conclude that in the minimal realization of the exotic Galilei
group the coordinate of the free particle cannot be commutative and
covariant simultaneously, cf. the case of the anyons \cite{CP}.
There exist at least three sorts of the coordinate, each of which
has definite advantages and disadvantages.

Duval and Horvathy showed \cite{DH} that within the minimal
realization, the coupling of the particle to  the arbitrary external
electric and magnetic fields can be achieved via  a simple
generalization of the free symplectic structure and Hamiltonian for
\begin{equation}\label{minEL}
    \sigma_{em}=dp_i\wedge dX_i +\frac{1}{2}\theta\epsilon_{ij}dp_i\wedge
    dp_j +\frac{1}{2}eB(X)\epsilon_{ij}dX_i\wedge dX_j,\qquad
    H_{em}=\frac{1}{2m}\vp\,{}^2+eV(X),
\end{equation}
where $V(X)$ is a scalar potential associated with the
electric field
$E_i=-\partial_i V(X)$. The Poisson brackets corresponding to the
$\sigma_{em}$ are
\begin{equation}\label{XXP}
    \{X_i,X_j\}=\frac{\theta}{1-e\theta B}\, \epsilon_{ij},\qquad
    \{X_i,p_j\}=\frac{1}{1-e\theta B}\, \delta_{ij},\qquad
    \{p_i,p_j\}=\frac{eB}{1-e\theta B}\, \epsilon_{ij},
\end{equation}
and the equations of motion for $X_i$ and $p_i$ take the form
similar to the $\theta=0$ case but with the mass $m$ changed for the
effective mass $m^*=m(1-e\theta B)$. The essential property of the
coordinate $Y_i$ defined by Eq. (\ref{Y}) is that it has the same
brackets (\ref{Yp}), (\ref{YYX}) in the presence of any magnetic
field $B(X)$ \cite{HPem}.

It is obvious that in the  case of the critical value of the
magnetic field $B=B_c\equiv(e\theta)^{-1}$, for which symplectic
form (\ref{minEL}) degenerates while brackets (\ref{XXP})
 blow up and the effective mass $m^*$ disappears,
 has to be treated separately \cite{DH,HPem}.  In \cite{HPem} it
was shown that in this case the system realizes a Hall-like motion,
which is described by the coordinate $Y_i$. On the other hand, it is
clear that in a generic case of the inhomogeneous magnetic field
there is a problem with realization of the operators satisfying the
quantum analogs of the Poisson bracket relations (\ref{XXP}).

The simultaneous commutativity and covariance of the coordinate can
be incorporated into the theory via the extended realization of the
exotic Galilei group \cite{HP2,HPem}. This is achieved by supplying
the phase space with the two additional canonically conjugate
translation-invariant variables $v_i$ associated with an
infinite-component Majorana-type representation of the exotic planar
Galilei group, being analogous to the Dirac $\alpha$ matrices. The
symplectic structure is given here by
\begin{equation}\label{sigma}
    \sigma=dp_i\wedge dx_i +\frac{1}{2}\kappa\epsilon_{ij}dv_i\wedge
    dv_j,
\end{equation}
and the rotation and the boost generators are realized in the form
\begin{equation}\label{JKex}
    {\cal J}=\epsilon_{ij}x_ip_j+\frac{1}{2}\kappa v_i^2,\qquad {\cal
    K}_i=mx_i-tp_i+\kappa\epsilon_{ij}v_j,
\end{equation}
while as before, the translation generator is identified with $p_i$.
Require that the first Casimir element from (\ref{Cas}) takes zero
value. Then, with taking into account (\ref{JKex}), we fix the form
of the Hamiltonian,
\begin{equation}\label{Hex}
    {\cal H}=\vp\, \vv -\frac{1}{2}m\vv\,{}^2,
\end{equation}
and find the equations of motion generated by it,
\begin{equation}\label{eqex}
    \dot{x}_i=v_i,\qquad
    \dot{p}_i=0,\qquad
    \dot{v}_i=\omega\epsilon_{ij}(v_j-m^{-1}p_j),
\end{equation}
where $\omega=m/\kappa$. Like in the case of the Dirac equation,
Hamiltonian (\ref{Hex}) is linear in momenta, the velocities are
noncommuting, $\{v_i,v_j\}=-\kappa^{-1}\epsilon_{ij}$, and in the
evolution of the covariant coordinate $x_i$, $\{x_i,x_j\}=0$, there
appears a Zitterbewegung-like term:
$$
x_i(t)=X_i(0)+\frac{1}{m}p_i t- \omega^{-1}\epsilon_{ij}V_j(t),
$$
where
\begin{equation}\label{Xex}
    X_i=x_i+\frac{\kappa}{m}\epsilon_{ij}V_j,
\end{equation}
\begin{equation}\label{V}
    V_i=v_i-m^{-1}p_i,
\end{equation}
and  $V_i(t)=(\cos\omega t\, \cdot \delta_{ij}+\sin\omega t \,\cdot
\epsilon_{ij})V_j(0)$. The quantities $V_i$ form a planar vector
invariant with respect to the space translations and boosts,
$\{{\cal K}_i,V_j\}=0$, $\{p_i,V_j\}=0$, and can be associated with
the internal rotation.

The quantity (\ref{Xex})
 has the same
transformation properties under the action of ${\cal P}_i$, ${\cal
K}_i$ and ${\cal J}$ as the coordinate $x_i$. Unlike the $x_i$, it
is Zitterbewegung-free, $\dot{X}_i=m^{-1}p_i$, and has the
non-commuting components, $\{X_i,X_j\}=\theta\epsilon_{ij}$ [cf. the
properties of the covariant coordinate $X_i$ within the minimal
realization]. The $X_i$  is analogous to the Foldy-Wouthuysen
coordinate for the Dirac particle. The combination ${\cal
X}_i=X_i-\frac{1}{2}\theta\epsilon_{ij}p_j$ (with $X_i$ given by
(\ref{Xex})) is also Zitterbewegung-free, it has commuting
components, but is not covariant under the action of the Galilei
boosts [cf. the properties of the coordinate (\ref{calX})]. It is
analogous to the Newton-Wigner coordinate for the Dirac particle
\cite{HP1}.

It is interesting to note that the dynamical picture of the extended
formulation turns out to be exactly the same as that for the usual
planar particle ($\theta=0$) subjected to the external homogeneous
magnetic and electric fields \cite{OP}.

The Hamiltonian and the rotation generator are represented
equivalently in the form
\begin{equation}\label{Hg}
    H=\frac{1}{2m}\vp\,{}^2-\frac{1}{2}m\vec{V}^2,
\end{equation}
$$
{\cal J}=\epsilon_{ij}X_ip_j+\frac{1}{2}\theta\vp\,{}^2+
\frac{1}{2}\kappa\vec{V}^2,
$$
while the boost generator takes the same form as in (\ref{Min}) with
$X_i$ given by Eq. (\ref{Xex}).
 We have not fixed yet the second Casimir element, which is
reduced here to the integral of motion associated with the
Zitterbewegung (circular motion), ${\cal C}_2=m^2\vec{V}^2$. Such a
Hamiltonian system corresponds to a special non-relativistic limit
applied to the model of relativistic particle with torsion
\cite{Ptor} associated with the (2+1)-dimensional analog of the
Majorana equation and underlying the theory of relativistic anyons
\cite{HP1}.
Like the relativistic analog, the present system is described by the
higher-derivative Lagrangian
\begin{equation}\label{Ltheta}
    L=\frac{1}{2}m\dot{x}_i^2
    +\theta\epsilon_{ij}\dot{x}_i\ddot{x}_j,
\end{equation}
which was analysed by Lukierski, Stichel and Zakrzewski \cite{LSZ}
(ignoring its relation to the relativistic higher-derivative model
\cite{Ptor}). In accordance with the Ostrogradski theory of
higher-derivative systems, at the Hamiltonian level the velocity
components $\dot{x}_i$ are identified as independent phase space
variables $v_i$.

{}From the structure of the Hamiltonian (\ref{Hg}) and equivalent
form of the symplectic structure (\ref{sigma}),
\begin{equation}\label{sigV}
    \sigma=dp_i\wedge dX_i+\frac{1}{2}\theta \epsilon_{ij}
    dp_i\wedge dp_j +\frac{\kappa}{2}\epsilon_{ij}dV_i\wedge dV_j,
\end{equation}
 it is clear that the system
(\ref{Ltheta}) describes not a free particle in the noncommutative
plane but a sort of rotator with degrees of freedom of the ghost
nature since they contribute a negative kinetic term into the
Hamiltonian. In order to reduce this system to a free exotic
particle of Duval and Horvathy \cite{DH} (which corresponds to a
minimal realization of the two-fold centrally extended Galilei
group), it is sufficient to fix the second Casimir element by
introducing the second class constraints $V_i=0$, $i=1,2$
\cite{HPem}. {}From the point of view of such a reduction, the
coordinate (\ref{Xex}) is the extension of the initial coordinate
$x_i$ commuting with the second class constraints \cite{OP}.

There is also another possibility to reduce the system
(\ref{Ltheta}), preserving the linear in the momentum Hamiltonian
structure (\ref{Hex}) similar to that of the Dirac equation. Instead
of the two second class constraints, the physical subspace of the
system can be singled out by imposing a complex polarization
condition given by one first class complex constraint
\begin{equation}\label{V-}
    V_-=0,
\end{equation}
$V_-=V_1-iV_2$. Then at the quantum level a state of the system can
be decomposed into the series in the Fock space states associated
with the velocity variables $\hat{v}_\pm=\hat{v}_1\pm i\hat{v}_2$,
$\vert\Psi\rangle=\sum_{k=0}^{\infty}\psi_k\vert k\rangle_v$, where
$\hat{v}_-\vert 0\rangle_v=0$. As a result, the quantum system will
be described by the pair of the infinite-component wave equations
\cite{HP2}
\begin{equation}\label{Sch}
    i\partial_t\psi_k+\sqrt{\frac{k+1}{2\theta}}\,\frac{\hat{p}_+}{m}\,\psi_{k+1}=0,
\end{equation}
\begin{equation}\label{Vquant}
    \hat{p}_-\psi_k+\sqrt{\frac{2(k+1)}{\theta}}\,\psi_{k+1}=0,
\end{equation}
where $k=0,1,\ldots,$ and $\hat{p}_\pm=\hat{p}_1\pm i\hat{p}_2$. Eq.
(\ref{Sch}) is the Schr\"odinger  equation corresponding to the
classical Hamiltonian (\ref{Hex}), while Eq. (\ref{Vquant}) is the
quantum analog of the classical constraint (\ref{V-}), whose role is
to separate effectively only one independent physical field degree
of freedom. The set (\ref{Sch}), (\ref{Vquant}) has the sense of the
infinite-component wave equations of the Dirac-Majorana-Levy-Leblond
type for the exotic particle, associated with the two-fold central
extension of the planar Galilei group. It was obtained in \cite{HP2}
by applying a special Jackiw-Nair non-relativistic limit \cite{JN}
to the spinor set of the equations proposed earlier in \cite{Any}
for the description of relativistic anyons.

Having in mind the discussed nature of the coordinates which appear
in the minimal realization of the exotic Galilei group, it is clear
that the coupling prescription (\ref{minEL}) in the case of the
Dirac theory would correspond to the minimal coupling in terms of
the Foldy-Wouhtuysen coordinates. Since the extended formulation of
a free exotic particle results in the free  wave equations
(\ref{Sch}), (\ref{Vquant}) realized in terms of the commuting
covariant coordinates $x_i$, it is natural to expect that the
coupling of the system to external electric and magnetic fields
proceeding from the extended formulation would be more close in
nature to the usual minimal coupling prescription of the Dirac
theory.

The coupling of the exotic particle to external electric and
magnetic fields in the extended formulation can be realized  as
follows \cite{HPem}. Modify the complex polarization condition
(\ref{V-}) via the minimal coupling prescription, $p_i\rightarrow
P_i=p_i-eA_i(x)$, $\epsilon_{ij}\partial_iA_j=B$.  Then the
generalization of the Hamiltonian (\ref{Hex}) can be fixed from the
requirement  of its (weak) commutativity with the changed
polarization condition. The essential feature of such a coupling
scheme is that the two real constraints
\begin{equation}\label{Lam}
    \Lambda_i=v_i-\frac{1}{m}P_i=0,\qquad
    \{\Lambda_i,\Lambda_j\}=-\kappa^{-1}(1-\beta)\epsilon_{ij},
\end{equation}
$\beta=\beta(x)=e\theta B(x)$, corresponding to one complex
polarization condition, change their nature from the second class
into the first class constraints at the critical value of the
magnetic field, $B=B_c$. As a result, at $B=B_c$, the constraints
(\ref{Lam}) eliminate not one but two degrees of freedom, leaving
only one degree described effectively  by the noncommutative
coordinate $Y_i$ \cite{HPem}. In a generic case, the classical
Hamiltonian weakly commuting with constraints (\ref{Lam}) and
reducing to the Hamiltonian (\ref{Hex}) in the free case, has the
form $\tilde{\cal H}=H_B+U$, with
\begin{equation}\label{HBV}
    H_B=\frac{1}{1-\beta}(P_i-\beta v_i)v_i-\frac{1}{2}mv_i^2,
\end{equation}
and $U$ being an arbitrary function of $X_i$, or $Y_i$.

In the case of homogeneous magnetic field different from the
critical one and for zero electric field  ($U=0$), the obtained
system describes the Landau problem  in the noncommutative plane. It
is necessary to distinguish the cases of subcritical and
overcritical magnetic fields. Assume that $e\theta>0$. Then the
physical states for $B<B_c$ are separated by the quantum
polarization condition
\begin{equation}\label{L-}
    \hat{\Lambda}_-\vert \Psi \rangle=0.
\end{equation}
The solutions of Eq. (\ref{L-}) describe the physical states of the
form
\begin{equation}\label{PhysSt}
    \vert \Psi\rangle_{phys}=\exp \left(\frac{1}{2}\theta
    m\hat{P}_-\hat{v}_+\right) \left(\vert 0\rangle_v \vert\psi\rangle\right),
\end{equation}
where $\vert 0\rangle_v$, $\hat{v}_-\vert 0\rangle_v=0$, is the
vacuum state of the Fock space generated by the velocity operators,
and $\vert\psi\rangle$ is a velocity-independent state associated
with other degrees of freedom.  The action of the Hamiltonian
operator corresponding to (\ref{HBV}) is reduced on the states
(\ref{PhysSt})
 to
\begin{equation}\label{Hphys}
     \hat{H}_B\vert \Psi\rangle_{phys}=\exp \left(\frac{1}{2}\theta
    m\hat{P}_-\hat{v}_+\right) \left(\vert 0\rangle_v
    \hat{H}_*\vert\psi\rangle\right),\qquad
    \hat{H}_*=\frac{1}{2m^*}\hat{P}_+\hat{P}_-.
\end{equation}
For $B<0$, the spectrum of the system is characterized by the energy
values  $E_N=e\vert B\vert N/m^*$, $N=0,1,\ldots$, and by the
angular momentum values $j=N,N-1,\ldots$. For $0<B<B_c$,
 $E_N=e\vert B\vert (N+1)/m^*$, $N=0,1,\ldots$, and
 $j=-N,-N+1,\ldots$ \cite{HPem}. The structure of the physical states
 is essentially different for $B<0$ and $0<B<B_c$:
in the former case, the finite number of the velocity Fock space
states $\vert n\rangle_v$, $n=0,\ldots, N$, contribute to a physical
state, while in the latter case all the infinite tower of the
velocity Fock states ($n=0,1,\ldots$) contributes to it. It is
essential, however, that in the both cases the common  eigenstates
of the energy and angular momentum are normalisable. In the critical
case, due to the first class nature of the constraints (\ref{Lam}),
equation(\ref{L-}) should be supplemented with the quantum condition
$\hat{\Lambda}_+\vert \Psi\rangle=0$. The solutions of these two equations
are given by the wave functions proposed by Laughlin to describe the
ground states in the fractional quantum Hall effect \cite{Laugh},
and coincide with the solutions of the equation (\ref{L-}) taken in
the limit $B\rightarrow B_c$, for the details see ref. \cite{HPem}.

In the case of overcritical magnetic field $B>B_c$ the solutions of
the quantum equation (\ref{L-}) are not normalisable \cite{HPem}.
The reason of this is rooted in a simple observation. In accordance
with Eq. (\ref{Lam}), the brackets between constraints $\Lambda_i$,
$i=1,2$,  for $B>B_c$ have an opposite sign in comparison with the
subcritical case $B<B_c$. It means that the operator
$\hat{\Lambda}_-$ being an annihilation-like operator for $B<B_c$,
transforms  into the creation-like operator having no nontrivial
kernel for $B>B_c$. Therefore, in the overcritical case, the
physical states have to be separated by the quantum condition
$\hat{\Lambda}_+\vert \Psi\rangle=0$ instead of the condition
(\ref{L-}). This change has to be accompanied by the  change of the
direction of time, $t\rightarrow -t$ \cite{HPem,Pet}.

It  was observed in  \cite{HPem} that in a generic case of
inhomogeneous magnetic field the quantum analog of the classical
Hamiltonian (\ref{HBV}) commuting with the quantum condition
(\ref{L-}) has a nonlocal nature. On the other hand, one notes that
there exists a class of the quantum systems with
coordinate-dependent mass related to some quasi-exactly solvable
systems \cite{Mx}. This, probably, indicates that for inhomogeneous
magnetic field of a special form the problem of non-locality of the
quantum Hamiltonian can be solved using some ideas related to
quasi-exact solvability and supersymmetry \cite{KP}.

Since the exotic particle system in the noncommutative plane is
related via a special non-relativistic limit to the relativistic
anyon, this means that the phenomenon similar to the existence of
the critical magnetic field should also exist if one couples the
latter system to the external electromagnetic field. The problem of
non-locality should also reveal itself there for electromagnetic
field of a generic form.

\vskip 0.2cm\noindent {\bf Acknowledgements}.
The work was supported
by the FONDECYT-Chile (project 1050001). I thank the organizers of
the SQS-05 Workshop (JINR, Dubna, Russia) and the Summer
Mini-Workshop in Theoretical Physics (2006)  (CECS, Valdivia, Chile)
for hospitality.



\end{document}